# A theoretical approach to study J/Ψ suppression in relativistic heavy ion collisions

**Santosh K. Karn**

**Department of Physics, School of Basic Sciences and Research, Sharda University, Greater Noida, NCR-Delhi, India.**

Email: skarn03@yahoo.com; santosh.karn@sharda.ac.in

## Abstract

With a view to understanding J/Ψ suppression in relativistic heavy ion collisions, we compute the suppression rate within the framework of hydrodynamical evolution model. For this, we consider an ellipsoidal flow and use an ansatz for temperature profile function which accounts for time and the three dimensional space evolution of the quark-gluon plasma. We have calculated the survival probability separately as the function of transverse and longitudinal momentum. We have shown that previous calculations are special cases of this model.

## 1. Introduction

Relativistic heavy ion collisions vis-à-vis formation of quark-gluon plasma (QGP) and its implications have been very excited field of research [1,2] and is going to remain very active for a long time[2]. For a detailed study, text books and review articles on the subject, we refer references given in [2]. In fact, the J/Ψ suppression has been theoretically argued [3] as an important signal for the formation of QGP in the collision experiments. As the science of small is deeply connected with the science of large, the study of J/Ψ suppression in the evolution of QGP in such experiments has many implications in astroparticle physics and cosmological evolution as well.

In the present work, we consider ellipsoidal evolution and use exponential type temperature profile function. In the next section, we briefly outline various theoretical works on the understanding of J/Ψ suppression. In section 3, we compute the suppression rate within the framework of the present model. Results are discussed and summarized in section 4.

## 2. Various theoretical approaches

The formation and evolution of QGP in the relativistic heavy ion collisions is well described within the framework of Bjorken hydrodynamical evolution model [4] in terms of thermodynamical variables, namely temperature (T), entropy density (s), energy density (ε), etc. The probability of disintegration is related to survival probability (S) which is given [5] by

$$P = 1 - S = 1 - \exp(-T) \tag{1}$$

In all the works on the J/Ψ suppression, survival probability has been calculated by parameterizing the thermodynamical quantities, namely the entropy density s [6], energy density ε [7], and temperature function T [8,9] respectively as

$$s(t_0, r) = s_0 (1 - r^2/R^2)^{a'} \; ; \; (r) = \varepsilon_0 (1 - r^2/R^2)^{2/3} \; ; \; T(r) = T(0)(1 - r^2/R^3)^{b/3} \; ; \; \text{and } T(r,t) = T_0 \exp(-br/R)\exp(-\eta z) t^{-1/3}$$

(2)

where R is the projectile radius, a ' is a free parameter[6], b is a parameter fixed as 1/3 [8], and  is a parameter which is a measure of steepness of the fall of temperature along z-direction [9]. In all the works except [9], the dependence of S on transverse momentum $p_T$ has been calculated in the non transparent manner while in [9] a cylindrical interaction volume is considered and J/Ψ suppression has been studied. However, in the present work, we make an attempt to consider an ellipsoidal interaction volume and compute the survival probability of J/Ψ suppression which is described in the next section.

## 3. Present model

In this section, we consider ellipsoidal evolution in both time and three dimensional space within the framework of hydrodynamical model for QGP. We use the concept of temperature density noramlized to unity over the plasma volume [10, 9] and compute the survival probability S as a function of $p_L$ and $p_T$ respectively.

For ellipsoidal symmetry, the four dimensional temperature profile function, in general, can be written as

$$T(t, \vec{r}') = T_0 f(t, z) T(\vec{r}) \tag{3}$$

where $f(t,z)$ is a function of time and z axis along the collision direction; $\vec{r}'$ is a function of x,y and z; and  $\vec{r}$ is a function of x and y in the transverse plane in the ellipsoidal interaction volume under consideration.  Here we have considered various functions as

$$f(t,z) = t^{-1/3} e^{-\eta z} \; ; T(\vec{r}) = e^{-\alpha_1 x} e^{-\alpha_2 y} ; and \; T(\vec{r}) f(t,z) = T(\vec{r}') \tag{4}$$

where $\eta$, $\alpha_1$ and $\alpha_2$ are measure of steepness of the fall of temperature along z, x and y directions. Here, $\alpha_1 = b_1/R_1$, $\alpha_2 = b_2/R_2$ with $R_1$ and $R_2$ are radius of colliding nuclei and $b_1$ and $b_2$ are parameters. The time dependence of the temperature profile function given in eq. (3) is in accordance with the scaling law [11]. We normalize the temperature profile function to unity over the ellipsoidal plasma volume as

$$\int T(t, \vec{r}') \, d\tau' = 1 \tag{5}$$

where $d\tau'$ is an ellipsoidal evolution volume element. Eqs. (3) and (4) lead to

$$T_0 \int t^{-1/3} \, dt \int e^{-\eta z} \, dz \int e^{-\alpha_1 x} \, dx \int e^{-\alpha_2 y} \, dy = 1 \tag{6}$$

Solving integrals in eq. (6), we obtain the normalization constant as

$$T_0 = (\eta \, \alpha_1 \alpha_2 / 12) \left[ \left( t_f^{2/3} - t_i^{2/3} \right) \sinh(\eta c) \cdot \sinh\left( \alpha_1 a \sqrt{1 - h^2/c^2} \right) \cdot \sinh\left( \alpha_2 b \sqrt{1 - h^2/c^2} \right) \right]^{-1} \tag{7}$$

where a, b and c are along x, y and z directions respectively; h within sine and cosine hyperbolic functions represents z co-ordinate such that x varies from $-a\sqrt{1 - h^2/c^2}$ to $+a\sqrt{1 - h^2/c^2}$ ; y varies from $-b\sqrt{1 - h^2/c^2}$ to $+b\sqrt{1 - h^2/c^2}$ and z varies from –c to +c. The plasma evolution ellipsoidal volume, $V = \int F(h) dh$ = (4/3) $\pi abc$ where the area F (h) = $\pi ab(1 - h^2/c^2)$.

Therefore, the temperature profile function becomes

$$T(t, \vec{r}') = (\eta \, \alpha_1 \alpha_2 / 12) \left[ \left( t_f^{2/3} - t_i^{2/3} \right) \sinh(\eta c) \cdot \sinh\left( \alpha_1 a \sqrt{1 - h^2/c^2} \right) \cdot \sinh\left( \alpha_2 b \sqrt{1 - h^2/c^2} \right) \right]^{-1} \cdot t^{-1/3} \, e^{-\eta z} \, e^{-(\alpha_1 x + \alpha_2 y)} \tag{8}$$

As momentum $\vec{p}$ is the conjugate of position vector $\vec{r}'$, we take the Fourier transforms of $T(t, \vec{r}')$ in eq. (3) with respect to $\vec{r}'$ which finally leads to $p_L$ and $p_T$ dependences of survival probability S. Therefore, we write

$$T(p_L, p_T) = \int T(t, \vec{r}') \, e^{i \, \vec{p} \cdot \vec{r}'} \, dt \, d\tau' \tag{9}$$

From eq. (9), we obtain

$$|T(p_L, p_T)| = (3/2) T_0 \left( t_f^{2/3} - t_i^{2/3} \right) |I_z| \, |I_x| \, |I_y| \tag{10}$$

where $|I_z| = \sqrt{(I_z I_z^*)}$, $|I_x| = \sqrt{(I_x I_x^*)}$ , $|I_y| = \sqrt{(I_y I_y^*)}$ with $I_z = \int e^{i p_L z} \, e^{-\eta z} \, dz$, $I_x = \int e^{i p_{Tx} x} \, e^{-\alpha_1 x} \, dx$, and $I_y = \int e^{i p_{Ty} y} \, e^{-\alpha_2 y} \, dy$.

Therefore, we obtain

$$|T(p_L, p_T)| = \frac{[\sinh^2 c \cos^2 p_L c + \cosh^2 c \sin^2 p_L c]^{1/2}}{\sinh c \cdot (^2 + p_L^2)^{1/2}} \cdot$$

$$\alpha_1 \frac{[\sinh^2 \alpha_1 a(1-z^2/c^2)^{\frac{1}{2}} \cos^2 p_{Tx} a(1-z^2/c^2)^{\frac{1}{2}} + \cosh^2 \alpha_1 a(1-z^2/c^2)^{\frac{1}{2}} \sin^2 p_{Tx} a(1-z^2/c^2)^{\frac{1}{2}}]^{1/2}}{(\alpha_1^2 + p_{Tx}^2)^{1/2} \sinh \alpha_1 a(1-z^2/c^2)^{\frac{1}{2}}} \cdot$$

$$\alpha_2 \frac{[\sinh^2 \alpha_2 b(1-z^2/c^2)^{\frac{1}{2}} \cos^2 p_{Ty} b(1-z^2/c^2)^{\frac{1}{2}} + \cosh^2 \alpha_2 b(1-z^2/c^2)^{\frac{1}{2}} \sin^2 p_{Ty} b(1-z^2/c^2)^{\frac{1}{2}}]^{1/2}}{(\alpha_2^2 + p_{Ty}^2)^{1/2} \sinh \alpha_2 b(1-z^2/c^2)^{\frac{1}{2}}}$$

(11)

From eq. (11), we obtain the value of $|T(P_T)|$ and $|T(P_L)|$. When $p_L \to 0$, we obtain an expression for $|T(P_T)|$ and the expression for $|T(P_L)|$ is obtain by taking $p_T \to 0$. The obtained expressions are

$$|T(p_T)| =$$

$$\alpha_1 \frac{[\sinh^2 \alpha_1 a(1-z^2/c^2)^{\frac{1}{2}} \cos^2 p_{Tx} a(1-z^2/c^2)^{\frac{1}{2}} + \cosh^2 \alpha_1 a(1-z^2/c^2)^{\frac{1}{2}} \sin^2 p_{Tx} a(1-z^2/c^2)^{\frac{1}{2}}]^{1/2}}{\sinh(\alpha_1 a(1-z^2/c^2)^{\frac{1}{2}}) \cdot (\alpha_1^2 + p_{Tx}^2)^{1/2}} \cdot$$

$$\alpha_2 \frac{[\sinh^2 \alpha_2 b(1-z^2/c^2)^{\frac{1}{2}} \cos^2 p_{Ty} b(1-z^2/c^2)^{\frac{1}{2}} + \cosh^2 \alpha_2 b(1-z^2/c^2)^{\frac{1}{2}} \sin^2 p_{Ty} b(1-z^2/c^2)^{\frac{1}{2}}]^{1/2}}{\sinh \alpha_2 b(1-z^2/c^2)^{\frac{1}{2}} \cdot (\alpha_2^2 + p_{Ty}^2)^{\frac{1}{2}}}$$

(12)

and

$$|T(p_L)| =$$

$$\frac{\eta}{\sinh \eta c \cdot (\eta^2 + p_L^2)^{1/2}} [\sin^2 hc \cos^2 p_L c + \cos^2 hc \sin^2 p_L c]^{1/2}$$

(13)

The expression for survival probability S of the J/Ψ suppression in the relativistic heavy ion collisions is obtained by substituting the value of $|T(P_T)|$ and $|T(P_L)|$ from eqs. (12) and (13) in eq. (1). It is to be noted here that eq. (5) ensures the dimensionless character of $|T(P_T)|$ and $|T(P_L)|$ given in eqs. (12) and (13) and subsequently that of S in eq. (1). It is seen that the expression for suppression rate of momentum along longitudinal direction in the ellipsoidal evolution remains the same as that in cylindrical evolution, i.e. eq. (13) is the same as eq. (6) in [9]. But the expression for suppression rate of momentum along transverse direction in the ellipsoidal evolution given by eq. (12) does not remain the same as obtained by others [6 - 9]. That is transverse momentum evolution is different in ellipsoidal flow than in other types of evolution considered, namely the spherical or, the cylindrical evolutions. However, the expression for suppression rate of momentum along transverse direction in the cylindrical and spherical evolutions can be reproduced simply by substituting a = b and a = b = c respectively in eq. (12). That is, the previous calculations are special cases of the present work. In order to understand the data of the J/Ψ suppression completely we argue that it is necessary to understand together the suppression rate of momentum along the longitudinal direction and the transverse direction. The total energy can be obtained by using eq. (3). In fact, by substituting the value of $T(t, \vec{r})$ from eq. (3) in the energy density equation $\varepsilon(t,r) = (\alpha_S \pi^2/15) \, T^4(t,r')$, one can obtain the expression for suppression rate of energy density along the longitudinal and transverse directions. The total energy can also be obtained from the equation, E = $((4/3) \, \pi abc. \varepsilon)$.

## 4. Discussion and summary

Within the framework of the Bjorken hydrodynamical model we have tried to understand the time and space evolution of QGP by considering ellipsoidal flow of the fluid and exponential fall of temperature in longitudinal and transverse directions. In the present work we have calculated the J/Ψ suppression in the relativistic heavy ion collisions not only of transverse momentum $P_T$ but also of longitudinal momentum $P_L$. Presently available data [12] are not sufficient for longitudinal momentum for comparison with our predicted results. Therefore further experimental works for extracting the information on this aspect is desirable. In fact, the study of the survival probability with respect to $P_T$ and $P_L$ dependences together will throw light for the complete understanding of the phenomenon of J/Ψ suppression. It is to note here that from the results thus obtained, one can also calculate the dependence of J/Ψ suppression with transverse and longitudinal energy density by using the relation between energy density and the corresponding temperature, and the results can be compared with the experimental data which we hope to address in the upcoming research works.

## Acknowledgements

The author is greatly benefited from discussion with Professor H Satz at T I F R and he wishes to express his gratitude to him. Author also wishes to thank Prof. R S Kaushal and Prof. Permanand for discussions. The author also wishes to thank the management of PCCS, Dr. APJ Abdul Kalam Technical University, India where the author worked for quite a long time and up to October 03, 2019 forenoon. Thanks are also due to the Head of the Department of Physics, Dean of the School of Basic Sciences and Research, and Dean, RTDC, Sharda University for providing the facilities to complete the work. The author also wishes to thank Prof. Ashok Kumar and Suvrat Karn for helping me in the typing work.